# Theoretical Analysis of Compressive Sensing via Random Filter

Lianlin Li,  Yin Xiang  and  Fang Li

Institute of Electronics, Chinese Academy of Sciences, Beijing, China, 100190

[lianlinli1980@gmail.com](lianlinli1980@gmail.com)

*Abstract:*

In this paper, the theoretical analysis of compressive sensing via random filter, firstly outlined by J. Romberg [compressive sensing by random convolution, *submitted to SIAM Journal on Imaging Science on July 9, 2008*], has been refined or generalized to the design of general random filter used for compressive sensing. This *universal* CS measurement consists of two parts: one is from the convolution of unknown signal with a random waveform followed by random time-domain subsampling; the other is from the directly time-domain subsampling of the unknown signal. It has been shown that the proposed approach is a universally efficient data acquisition strategy, which means that the *n*-dimensional signal which is *S* sparse in any sparse representation can be exactly recovered from $m \geq S\log n$ measurements with "overwhelming" probability.

I.  INTRODUCTION

Recently, a new emerging field has made a paradigmatic step in the way information is presented, stored, transmitted and recovered. This area is often referred to as compressive sensing (or compressed sensing, compressed sampling, etc) developed by Donoho, Tao, Candes and Romberg et al [1-4]. The common approaches to sample a signal generally follow the well-known Nyquist-Shannon's theorem: the sampling rate must be at least twice the maximum frequency present in the signal. Consequently, before the implement of standard analog-to-digital converter, one must ensure that (1) the present signal must be (or approximate) compressible in the finite frequency domain , i.e. the bandlimited signal; (2) the base-band conversion aimed at making the signal centered at zero-frequency is carried out and then (3) an anti-aliasing low-pass filter to band-limit the signal is employed. Finally, the signal is uniformly sampled at or above the Nyquist rate. Besides the expensive cost for realizing the above operations, the practical difficulties due to the unknown (or large) signal bandwidth often exists (even non-overcomeable in some cases) in the community of wireless communication [5], ultra-wide radar imaging, beamforming, and so on.

As we know, the Nyquist rate is a sufficient, by no means necessary condition. Only the *prior* information, the signal bandwidth or approximate bandwidth is used for the signal sampling based on the Nyquist-Shannon theorem. In practice, the signal or image is compressible in some basis, which means the information of signal can be captured by much smaller number of coefficients than the length of the signal/image. The CS theory asserts that one can recover certain signal or image from far fewer samples or measurements than traditional methods required when the signal of interest is compressible or sparse in some basis. The sparsity of signals is a fact often exploited in signal processing. In particular, the common way to compress a signal is to transform it to the basis in which it is sparse and subsequently store only the locations and values of the few non-zero elements. Recently, it has been discovered that, in addition to storage, the signal sparsity can be leveraged to reduce the number of measurements for signal acquisition and detection; it has been shown that, if a signal is sufficiently sparse, a small number of projections onto the random vectors is enough to recover the signal.

The CS measurements, different than samples that traditional analogy-to-digital converters take, model the acquisition of signal $x_0$ as a series of inner products against different the independent waveforms $\{\phi_k : k = 1, 2, 3, \cdots, m\}$ (from the discussion later, this paper deals with the case of $\mathrm{E}\langle \phi_i, \phi_j \rangle = 0$ if $i \neq j$, where the symbols "E" stands for the mean operation of some random variable), in particular,

$$y_k = \langle \phi_k, x_0 \rangle, \quad k = 1, 2, 3, \cdots, m \tag{1.1}$$

As well known, the recovering $x_0$ from $y_k$, a kind of classical linear inverse problem will need more measurements than unknowns, i.e. $m \geq n$. But the CS theory tell us that if the signal of interest $x_0$ is S-sparse in the orthogonal framework of $\Psi$ and the $\phi_k$ are chosen appropriately, then results from CS have shown us that recovering $x_0$ is possible even when there are far fewer measurements than unknowns, $m \ll n$. We say $x_0$ is S-sparse in $\Psi$ if we can decompose $x_0$ as $x_0 = \Psi \alpha_0$, where $\alpha_0$ has at most $S$ non-zero components. In some applications, the signals of interest are not perfectly sparse; however, all most of information can be captured by small number of terms. That is, there is a transform vector $\alpha_{0,S}$ with only S terms such that $\|\alpha_{0,S} - \alpha_0\|_2$ is small. This paper will focus on the CS recovery via $l_1$ minimization. Given the measurements $y = \Phi x_0$, we solve the convex optimization program

$$\min_{\alpha} \|a\|_{l_1} \quad \text{subject to} \quad y = \Phi \Psi \alpha \tag{1.2}$$

In words, (1.2) searches for the set of transform coefficients $\alpha$ such that the measurements of the corresponding signal $\Psi \alpha$ agree with $y$. The $l_1$-norm is being used to measure the sparsity of candidate signals.

In compressed sensing, the use of randomly generated projections to make measurements can sidestep the computational difficult task of checking whether the measurements can ensure the signal recovery. By considering recovery stochastically, it has been shown that measurements generated from Gaussian or Bernoulli random variables can ensure the signal recovery with high probability. But, these CS measurements can not be usually used in practice (at least can not used for the real-time purpose) because of its time-consuming computation and the difficulty of physical realization. In [7], Ailon and Chazelle has proposed the idea of randomized Fourier transform followed by a random projection as a fast Johnson-Lindenstrauss transform. They have showed that this matrix behave like a random waveform matrix with extremely high probability. Vertterli et al has developed alternative approach named as sampling signal with finite rate of innovation. The center idea is that the sampling rate for a sparse signal can be significantly reduced by first convolving with a kernel that spread it out [8]. In [6], the numerical simulations are carried out to demonstrate the recovery of sparse signals from a small number of samples of the output of a finite length "random filter". In [9] Romberg has developed a universal CS measurements by using a special random convolution (its amplitude of frequency-domain identically equal to 1) and derived bounds on the number of samples need to guarantee sparse reconstruction from a more theoretical perspective. Moreover, in [10] Bajwa et al has proposed the Toeplitz-structured compressed sensing matrices, which is not universal CS measurement and whose entries comes from the Bernoulli distribution. As a matter of fact, the *universal* CS measurement matrix proposed by Romberg belongs to a kind of special circulant Toeplitz matrix.

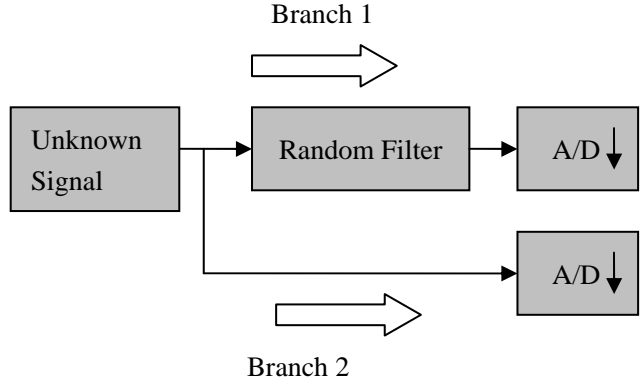

Fig.1 The proposed CS measurement

Inspired by the Romberg's results [10] and the idea of joint sparse representation of signal [11], a *universal* CS measurement has also been designed in this paper and consists of two branches (see Fig.1): one is used for the convolution of unknown signal with a random waveform followed by random time-domain subsampling; the other is from the directly time-domain subsampling of the unknown signal. As we known though the CS measurement belongs to lossy compress, the part of information to be sensed should be captured. Assume $x$ is a cell of signal (or a resolution unit of image) and we will sense it. CS will not check out information cell or resolution unit if all the information of this signal cell to be sensed is completely lost. Usually, joint measurement will be useful approach. As a matter of fact, the branch 2 is used to avoid this complete lossy of frequency-domain information of signal. It is noted that for two branches we "compress" the measurements by subsampling, in particular, we simply observe the entries of $H\alpha$ and $I\alpha$ at a small number of randomly chosen locations and throw the rest away. Convolving with the response of random filter $h$ can "compress" the information of $x$ into each sample of $Hx$ which can be untangled by solving (1.2). In this paper we employ the mathematical model for samples at random location, which means that to generate and i.i.d. sequence of Bernoulli random variables, each of which takes a value of 1 with probability $m/n$, and sample at locations $t$. Along the line proposed by Romberg, the theoretical analysis of sampling bound has been carried out.

## II.     MAIN RESLTS

Refer to Fig.1; our CS measurement process consists of two branches: (a) the branch 1 has two steps in which we let the signal $x_0 \in \mathbb{R}^n$ through a random filter described by $h \in \mathbb{R}^n$ and then subsample; (b) subsample directly the signal $x_0 \in \mathbb{R}^n$. The time-domain response of random filter $h \in \mathbb{R}^n$ is generated by *arbitrary* real-valued random distribution with zero mean, for example, the Gaussian distribution, the uniform distribution, the Bernoulli distribution, and so on. It can be found that the proposed CS matrix can be reduced to one proposed by Romberg and Bajwa et al. In terms of linear algebra, we can write the time-domain pulse at the output of branch 1 as the form of convolution of $x_0$ and $h$, in particular, $Hx_0$, where $H = n^{-1/2} F^* \Sigma F$ with $F$ as the discrete Fourier matrix defined by

$$F_{t,\omega} = \exp\left(-j2\pi \frac{(t-1)(\omega-1)}{n}\right), \quad 1 \le t, \omega \le n$$

and

$$\Sigma = \begin{bmatrix} \tilde{\sigma}_1 & 0 & \cdots & \\ 0 & \tilde{\sigma}_2 & & \\ \vdots & & \ddots & \\ & & & \tilde{\sigma}_n \end{bmatrix}$$

where $\{\tilde{\sigma}(\omega)\}$ is the Fourier transform of the real-valued random variable $\{\sigma(t)\}$

$$\tilde{\sigma}(\omega) = \sum_{t=1}^{n} \sigma(t) \exp\left(-j2\pi \frac{(t-1)(\omega-1)}{n}\right) \qquad (2.1)$$

From the definition of FFT, one has

$$\operatorname{Im} \tilde{\sigma}(\omega_1) = \operatorname{Im} \tilde{\sigma}(\omega_{n/2+1}) = 0$$

and

$$\tilde{\sigma}_{\omega = \frac{n}{2}+2:n} = \tilde{\sigma}^*_{n-\omega+2} \text{ for } \omega = 2, 3, \cdots, n/2.$$

Taking as example (it can be generalized to general case along the almost identical line), we consider $\{\sigma(t)\}$ with i.i.d. Gaussian distribution with zero-mean and $1/n$ variance, i.e.

$$\{\sigma(t)\} \sim \frac{1}{\sqrt{n}} N(0,1) \qquad (2.2)$$

Now, the proposed CS measurement matrix can be rewritten as

$$H_c = \begin{bmatrix} H \\ \sqrt{n}I \end{bmatrix} \tag{2.3}$$

Unlike the CS measurement proposed by Romberg et al in which $H_c^* H_c = nI$, the proposed CS measurement matrix satisfies

$$\mathrm{E}(H_c^* H_c) = nI \tag{2.4}$$

because of $\mathrm{E}(H^* H) = n^{-1}\mathrm{E}(F^*\Sigma^* FF^*\Sigma F) = nI$ and $\mathrm{E}(H^* I) = \mathrm{E}(IH) = 0$ (In fact, the CS measurement matrix whose entries are generated by i.i.d. random variables also obeys this rule provided by (2.4)).

From above discussions, one has the following conclusions:

(1) $\langle \tilde{\sigma}(\omega_1)\tilde{\sigma}^*(\omega_2) \rangle = 0$ for $\omega_2 \neq \omega_1$; however $\langle |\tilde{\sigma}(\omega)|^2 \rangle = 1$ if $\omega_2 = \omega_1$

(2) $\langle \tilde{\sigma}_R(\omega) \rangle = \langle \tilde{\sigma}_I(\omega) \rangle = 0$ and $\langle |\tilde{\sigma}_R(\omega)|^2 \rangle = \langle |\tilde{\sigma}_I(\omega)|^2 \rangle = \dfrac{1}{2}$ or $\langle \tilde{\sigma}_R(\omega)\tilde{\sigma}_I(\omega) \rangle = 0$

where

$$\tilde{\sigma}_R(\omega) = \sum_{t=1}^{n} \sigma(t) \cos\left(2\pi \frac{(t-1)(\omega-1)}{n}\right)$$

and

$$\tilde{\sigma}_I(\omega) = -\sum_{t=1}^{n} \sigma(t) \sin\left(2\pi \frac{(t-1)(\omega-1)}{n}\right)$$

(3) $\langle \tilde{\sigma}_R(\omega_1)\tilde{\sigma}_R(\omega_2) \rangle = 0$, $\langle \tilde{\sigma}_I(\omega_1)\tilde{\sigma}_I(\omega_2) \rangle = 0$ and $\langle \tilde{\sigma}_R(\omega_1)\tilde{\sigma}_I(\omega_2) \rangle = 0$ if $\omega_2 \neq \omega_1$.

Furthermore, let us consider the entries of $H$ denoted by

$$\begin{aligned}
(H)_{i,j} &= \sum_{k=1}^{n} \tilde{\sigma}(\omega_k) f_{k,i}^* f_{k,j} \\
&= \tilde{\sigma}(\omega_1) f_{1,i}^* f_{1,j} + \tilde{\sigma}(\omega_{n/2+1}) f_{n/2+1,i}^* f_{n/2+1,j} + 2\sum_{k=2}^{n/2} \mathrm{Re}\left(\tilde{\sigma}(\omega_k) f_{k,i}^* f_{k,j}\right)
\end{aligned} \tag{2.5}$$

From (2,5), readily one has

$$\begin{aligned}
\sqrt{n}\,\text{Im}(H)_{i,j} &= \tilde{\sigma}_R(\omega_1)\text{Im}(f_{1,i}^* f_{1,j}) + \tilde{\sigma}_I(\omega_1)\text{Re}(f_{1,i}^* f_{1,j}) \\
&\quad + \tilde{\sigma}_R(\omega_{n/2+1})\text{Im}(f_{n/2+1,i}^* f_{n/2+1,j}) + \tilde{\sigma}_I(\omega_{n/2+1})\text{Re}(f_{n/2+1,i}^* f_{n/2+1,j}) \\
&= \tilde{\sigma}_I(\omega_1) + \tilde{\sigma}_I(\omega_{n/2+1})\text{Re}(f_{n/2+1,i}^* f_{n/2+1,j}) \\
&= 0
\end{aligned}$$

(2.6)

and

$$\begin{aligned}
\sqrt{n}\,\text{Re}(H)_{i,j} &= \tilde{\sigma}_R(\omega_1)\text{Re}(f_{1,i}^* f_{1,j}) - \tilde{\sigma}_I(\omega_1)\text{Im}(f_{1,i}^* f_{1,j}) \\
&\quad + \tilde{\sigma}_R(\omega_{n/2+1})\text{Re}(f_{n/2+1,i}^* f_{n/2+1,j}) - \tilde{\sigma}_I(\omega_{n/2+1})\text{Im}(f_{n/2+1,i}^* f_{n/2+1,j}) \\
&\quad + 2\sum_{k=2}^{n/2}\left\{\tilde{\sigma}_R(\omega_k)\text{Re}(f_{k,i}^* f_{k,j}) - \tilde{\sigma}_I(\omega_k)\text{Im}(f_{k,i}^* f_{k,j})\right\} \\
&= \tilde{\sigma}_R(\omega_1) + \tilde{\sigma}_R(\omega_{n/2+1})\text{Re}(f_{n/2+1,i}^* f_{n/2+1,j}) \\
&\quad + 2\sum_{k=2}^{n/2}\left\{\tilde{\sigma}_R(\omega_k)\text{Re}(f_{k,i}^* f_{k,j}) - \tilde{\sigma}_I(\omega_k)\text{Im}(f_{k,i}^* f_{k,j})\right\} \\
&= \tilde{\sigma}_R(\omega_1) + \tilde{\sigma}_R(\omega_{n/2+1})\cos(\pi(i-j)) \\
&\quad + 2\sum_{k=2}^{n/2}\left\{\tilde{\sigma}_R(\omega_k)\cos\left(\frac{2\pi}{n}(k-1)(i-j)\right) - \tilde{\sigma}_I(\omega_k)\sin\left(\frac{2\pi}{n}(k-1)(i-j)\right)\right\}
\end{aligned}$$

(2.7)

From (2.5) (2.6) and (2.7) one can find that (*a*) the entries of *H* are real which is reasonable because of real-valued system, and (*b*) *H* is circulant toeplitz-kind matrix in the form of

$$H = \begin{bmatrix} a_1 & a_2 & \cdots & a_{n-1} & a_n \\ a_n & a_1 & a_2 & \cdots & a_{n-1} \\ \vdots & a_n & a_1 & \cdots & a_{n-2} \\ \vdots & \vdots & \ddots & \ddots & \vdots \\ a_2 & \cdots & \cdots & a_n & a_1 \end{bmatrix}$$

(2.8)

Moreover, it is noted that

$$n\text{E}\{a_j a_{j'}\} = 1/2 + 1/2\cos(\pi(j-j')) + 2\sum_{k=2}^{n/2}\left\{\cos\left(\frac{2\pi}{n}(k-1)(j-j')\right)\right\}$$

(2.9)

From (2.9) it can be found that $\text{E}\{a_j a_{j'}\} \approx 1 - 1/n$ if $j = j'$; however $|\text{E}\{a_j a_{j'}\}| \leq 1/n$ if

$j \neq j'$, which means that $\{a_k\}$ can be looked as some variables generated by the i.i.d. $N(0,1)$, especially for larger *n*. It should be pointed that for CS measurement proposed by Romberg, it is also in the form of (2.8); however $\{a_k\}$ is *also* the same as description by (2.8) and (2.9).

Our theoretical result shows that if we generated pulse as above, then with high probability we will be above to sense the vase majority of signals supported on a fixed set in the $\Psi$ domain. This result can be summarized as following theorem:

**THEOREM 2.1.**

*Let $\Psi$ be an arbitrary signal representation. Fix a fixed $\Gamma$ of size $|\Gamma| = S$ in the $\Psi$ domain, and choose a sign sequence z on $\Gamma$ uniformly at random. Let $\alpha_0$ be a set of $\Psi$ domain coefficients supported on $\Gamma$ with sign z, and take $x_0 = \Psi \alpha_0$ as the signal to be acquired. Create a CS measurement matrix $H_c$ as described above, and choose a set of sample locations $\Omega$ of size $|\Omega| = m$ uniformly at random with*

$$m \geq C_0 S \log(n/\delta)$$

*and also $m \geq C_0' \log^3(n/\delta)$ where $C_0$ and $C_0'$ are known constants. Then given the set of samples on $\Omega$ of $H_c x_0$, the program (1.2) will recover $\alpha_0$ (hence $x_0$) exactly with probability exceeding $1 - O(\delta)$*

Roughly, theorem 2.1 works because the generated CS measurement matrix *H* will be incoherent with any fixed orthonormal matrix with 'overwhelming' probability. In section III, the detailed proofs will be provided.

## III. PROOFS

### 3.1 PREMILARY

In [3] Candes and Tao have show that if a measurement satisfying the so-called UUP (the uniform uncertainty principle) and ERP (the exact reconstruction principle), a *S*-sparse signal can be exactly reconstructed with whelming probability by solving (1.2). As a slight generalization of leamma 2.1 in [3], we give the following lemma.

**LEMMA 3.1**

*Assume that the measurement matrix $F_\Omega$ obeys ERP. We let f be a fixed signal of the form $f = f_0 + h$ where $f_0$ is a signal supported on a set T. Then with whelming probability, any $l_1$-minimizer obeys*

$$\|f^\#\|_{T^c, l_1} \leq \frac{2}{1-\alpha}\|h\|_{l_1} \qquad (3.1)$$

It is noted that in this paper the third condition of ERP is generalized into

$$P(t) \leq \alpha < 1 \quad \text{for all} \quad t \in T^c \qquad (3.2)$$

*Proof:*

Observe that since f is feasible for (2.1), we immediately have

$$\|f^\#\|_{l_1} \leq \|f\|_{l_1} \leq \|f_0\|_{l_1} + \|h\|_{l_1} \qquad (3.3)$$

Because ERP holds, one construct a function $P = F_\Omega^* V$ for some $V \in l_2(K)$ such that $P = \text{sgn}(f_0)$ on T and $|P(t)| \leq \alpha < 1$ on $t \in T^c$. Now one has the identity

$$\langle f^\#, P \rangle = \langle f_0 + h, P \rangle \qquad (3.4)$$

Consequently,

$$\langle f^\#, P \rangle = \langle f_0, P \rangle + \langle h, P \rangle \geq \|f_0\|_{l_1} - \|h\|_{l_1} \qquad (3.5)$$

On the other hand, the bounds on P give

$$|\langle f^\#, P \rangle| \leq \sum_T |f^\#| + \alpha \sum_{T^c} |f^\#| = \sum_{\mathbb{Z}_n} |f^\#| - (1-\alpha)\sum_{T^c} |f^\#| \qquad (3.6)$$

To conclude, we established that

$$\|f^\#\|_{T^c, l_1} \leq \frac{2}{1-\alpha}\|h\|_{l_1} \qquad (3.7)$$

∎

The Talagrand inequality is a key issue used for the proofs of our results, which is summarized by lemma 3.2.

**LEMMA 3.2** The Talagrand inequality

*Assume that $|f| \leq B$ for every f in F, and $Ef(Y_i) = 0$ for every f in F and i=1,2,3,...,n.*

Lets $Z = \sup_{f \in F} \sum_{i=1}^{n} f(Y_i)$, $\bar{Z} = \sup_{f \in F} \left| \sum_{i=1}^{n} f(Y_i) \right|$ and $\{Y_i\}_{i=1,2,\ldots,N}$ is the independent random variable from Banach space.

Then for all $t \geq 0$,

$$\Pr(|Z - EZ| > t) \leq 3 \exp\left(-\frac{t}{KB} \log\left(1 + \frac{Bt}{\sigma^2 + BE\bar{Z}}\right)\right) \tag{3.8},$$

where $\sigma^2 = \sup_{f \in F} \sum_{i=1}^{n} E f^2(Y_i)$, $K$ is a numerical constant.

## 3.2. COHERENCE BOUNDS

**LEMMA 3.3.**

*Let $\Psi$ be an arbitrary fixed matrix with $\|\psi_k\|_2 = 1$ for any $k$, and create $H$ at random as above with $H = n^{-1/2} F^H \Sigma F$. Choose $0 < \delta < 1$. Then with probability exceeding $1 - \delta$, the coherence $\mu(H, \Psi)$ will obey*

$$\mu(H, \Psi) \leq \sqrt{2 \log\left(\sqrt{\frac{2}{\pi}} \frac{n}{\delta}\right)}. \tag{3.9}$$

*Proof*: The proof is a simple application of properties of the normal distribution $N(0,1)$. According to above discussion, the entries of $H$ can be looked as one generated by $N(0,1)$, then $(H \cdot \Psi)_{i,j}$ can rewritten as

$$y = (H \cdot \Psi)_{i,j} = \sum_k h^k \psi_k \tag{3.10}$$

Due to $\|\psi_k\|_2 = 1$, one has the important conclusion as $y \sim N(0,1)$. After some simple manipulation, one has

$$\Pr(|y| \geq \lambda) \leq \sqrt{\frac{2}{\pi}} \exp\left(-\frac{\lambda^2}{2}\right) \tag{3.11}$$

Taking $\lambda^2 = 2 \log\left(\sqrt{\frac{2}{\pi}} \frac{n}{\delta}\right)$ and applying the union bound over all $n$ choices of all entries of $(H \cdot \Psi)_{i,j}$ established the lemma.

**LEMMA 3.4**

*Fix an arbitrary fixed matrix $\Psi$ with $\|\psi_k\|_2 = 1$ for any k, and a subset of the $\Psi$-domain $\Gamma = \{\gamma_1, \gamma_2, \cdots, \gamma_S\}$ of size $|\Gamma| = S$. Generate a CS measurement matrix H as described as Lemma 3.3, and let $r^k$ (k=1,2,...,n) be the rows of $H\Psi_\Gamma$. Then with probability exceeding $1-\delta$,*

$$v(\Gamma) := \max_{k=1,2,\cdots,n} \|r^k\|_2 \leq \sqrt{8S} \quad \text{for} \quad S \geq C\log(n/\delta),$$

*where C is a numerical constant.*

*Proof*: From lemma 3.3, it can found that any entries of $(H \cdot \Psi)_{i,j}$ obeys the normal distribution, i.e. $(H \cdot \Psi)_{i,j} \sim N(0,1)$. Consequently, $\|r^k\|_2^2$ obeys the $\chi^2$-distribution with freedom S, in particular,

$$\|r^k\|_2^2 = \sum_{j=1}^{S} (r_j^k)^2 \sim \chi^2(S) \tag{3.12}$$

If the random variable y obeys the $\chi^2$-distribution with freedom $n$, its function of probability density is given by

$$f_n(y) = \begin{cases} \dfrac{1}{2\Gamma(n/2)} \left(\dfrac{y}{2}\right)^{n/2-1} \exp(-y/2) & y > 0 \\ 0 & y \leq 0 \end{cases}$$

(3.13)

where $\Gamma(a) := \int_0^\infty x^{a-1} \exp(-x) dx$

To obtain the bound of $\|r^k\|_2^2$, we carried out the following operation

$$\Pr\left(\|r^k\|_2^2 \geq A\right) = \int_A^\infty f_S(y) dy$$

$$= \frac{1}{\Gamma(S/2)} \int_{A/2}^\infty t^{S/2-1} \exp(-t) dt$$

$$\leq \frac{1}{\Gamma(S/2)} (A/2)^{S/2-0.9} \exp(-A/2)$$

$$\tag{3.14}$$

The inequality is justified for $\|r^k\|_2^2 \geq 8S$ with probability exceeding $1-\delta$ when $S \geq C\log(n/\delta)$, where $C$ is a numerical constant.

## 3.3 SPARSE RECOVERY

The following results extends the main results of [2] and [9] to take advantage of our refined bound on the norm coherence. The following theorems are stated for general measurement system U with $\mathrm{E}(U^*U) = nI$. The proofs follow the same general outline put forth in [2] [9], with one important conditions for the successful recovery of a vector $\alpha_0$ supported on $\Gamma$ with sign sequence $z$ are that $\Phi_\Gamma$ has full rank, where $\Phi_\Gamma$ is the $m \times S$ matrix consisting of the columns of $\Phi$ indexed by $\Gamma$ and that

$$|\pi(\gamma)| = \left|\left\langle \left(\Phi_\Gamma^* \Phi_\Gamma\right)^{-1} \Phi_\Gamma^* \varphi_\gamma, z \right\rangle\right| < \alpha < 1, \text{ for all } \gamma \in \Gamma^c \tag{3.15}$$

where $\varphi_\gamma$ is the column of $\Phi$ at index $\gamma$. There are two essential steps in establishing theorem 2.1:

(1) Show that with probability exceeding, the random matrix will have a bounded inverse:

$$\left\|\left(\Phi_\Gamma^* \Phi_\Gamma\right)^{-1}\right\| \leq 2/m \tag{3.16}$$

where $\|\cdot\|$ is the standard operator norm. This step can be finished by theorem 3.

(2) Establish, again with probability exceeding $1-\delta$, that

$$|\pi(\gamma)| = \left|\left\langle \left(\Phi_\Gamma^* \Phi_\Gamma\right)^{-1} \Phi_\Gamma^* \varphi_\gamma, z \right\rangle\right| < \alpha < 1, \text{ for all } \gamma \in \Gamma^c$$

**THEOREM 3.1:**

*Let U be a matrix with* $\mathrm{E}(U^*U) = nI$. *Consider a fixed set T and let* $\Omega$ *be a random set sampled using the Bernoulli model. Then*

$$E\left\|\frac{1}{m}U_{\Omega T}^* U_{\Omega T} - I\right\| \leq C_R \frac{\sqrt{\log|T|}}{\sqrt{m}} v(\Gamma) \tag{3.17}$$

*for some positive constant* $C_R$, *provided the right-hand side is less than 1.*

*Proof:*

Introduce $Y = \frac{1}{m}\sum_{k=1}^{n}\delta_k u_k \otimes u_k - I$, obviously $EY = 0$. Along the line done by Candes et al, a symmetrization technique is also employed to derive the bound of the expected value of the norm of Y. To end this let Y' be an independent copy of Y, i.e.

$$Y' = \frac{1}{m}\sum_{k=1}^{n}\delta'_k u_k \otimes u_k - I \tag{3.18}$$

where $\{\delta'_k\}$ are independent copies of $\{\delta_k\}$, and write

$$E_\delta \|Y\| \leq E_{\delta,\delta'} \|Y - Y'\|$$

Now let $\{\varepsilon_k\}$ be a sequence of Bernoulli variables taking values $\pm 1$ with equal probability (and independent of the sequences $\{\delta'_k\}$ and $\{\delta_k\}$). We have

$$\begin{aligned} E_\delta \|Y\| &\leq E_{\delta,\delta'} \left\| \frac{1}{m}\sum_{k=1}^{n}(\delta_k - \delta'_k) u_k \otimes u_k \right\| \\ &\leq 2 E_{\delta,\varepsilon} \left\| \frac{1}{m}\sum_{k=1}^{n}\delta_k \varepsilon_k u_k \otimes u_k \right\| \end{aligned} \tag{3.19}$$

From the well-known Rudelson's theorem, i.e.

$$E_\varepsilon \left\| \sum_{k=1}^{n}\delta_k \varepsilon_k u_k \otimes u_k \right\| \leq \frac{C_R}{4}\sqrt{\log|T|} \cdot \sqrt{\left\|\sum_{k=1}^{n}\delta_k u_k \otimes u_k\right\|} \cdot \max_{k:\delta_k=1}\|u^k\|$$

We have

$$\begin{aligned} E_\delta \|Y\| &\leq \frac{C_R}{2}\frac{\sqrt{\log|T|}}{m} \cdot E_\delta \sqrt{\left\|\sum_{k=1}^{n}\delta_k u_k \otimes u_k\right\|} \cdot \max_{1\leq k\leq n}\|u^k\| \\ &\leq \frac{C_R}{2}\frac{\sqrt{\log|T|}}{m} \cdot \sqrt{E_\delta \left\|\sum_{k=1}^{n}\delta_k u_k \otimes u_k\right\|} \cdot \max_{1\leq k\leq n}\|u^k\| \end{aligned}$$

$$(3.20)$$

If $\frac{C_R}{2}\frac{\sqrt{\log|T|}}{m}\max_{1\leq k\leq n}\|u^k\| < 1$, above equations yields to

$$E_\delta \|Y\| \leq C_R \frac{\sqrt{\log|T|}}{m}\max_{1\leq k\leq n}\|u^k\| \tag{3.21}$$

Readily, the following conclusion exists:

$$E\|Y\| \leq 2a, \tag{3.22}$$

with

$$a = \frac{C_R}{2}\frac{\sqrt{\log|T|}}{m}\max_{1\le k\le n}\|u^k\| = \frac{C_R}{2}\frac{\sqrt{\log|T|}}{m}v(\Gamma). \qquad (3.23)$$

which concludes our proof of this theorem.

■

From Lemma 4 one has $a \le C_R \dfrac{\sqrt{2T\log|T|}}{m}$ with probability exceeding $1-\delta$.

With above theorem established, we have a bound on the expected value of $\left\|\dfrac{1}{m}U^*_{\Omega T}U_{\Omega T} - I\right\|$.

The following theorem shows that $\dfrac{1}{m}U^*_{\Omega T}U_{\Omega T}$ is close to the identity with high probability, turning the statement about expectation into a corresponding large deviation result.

**THEOREM 3.2**:

*Let U, T and $\Omega$ be as in above theorem. Suppose that the number of measurements m obeys*

$$m \ge \max\left(C_1 v\sqrt{\log|T|},\; C_2 v^2 \log(3/\delta)\right) \qquad (3.24)$$

*For some positive constants $C_1$, $C_2$. Then*

$$\Pr\left(\left\|\frac{1}{m}U^*_{\Omega T}U_{\Omega T} - I\right\| \ge \frac{1}{2}\right) \le \delta \qquad (3.25)$$

*where $\|\cdot\|$ is the standard operator $l_2$-norm here, the largest eignvalue (in absolute value).*

*Proof*: This proof is a straightforward application of the Talagrand inequality. Set

$$\begin{aligned}
Y &= \frac{1}{m}\sum_{k=1}^{n}\delta_k u_k \otimes u_k - I \\
&= \frac{1}{m}\sum_{k=1}^{n}\delta_k u_k \otimes u_k - \frac{1}{n}\sum_{k=1}^{n}\mathrm{E} u_k \otimes u_k \\
&= \sum_{k=1}^{n} Y_k
\end{aligned}$$

where $Y_k := \dfrac{1}{m}\delta_k u_k \otimes u_k - \dfrac{1}{m}\mathrm{E}\delta_k u_k \otimes u_k$. Note that $\mathrm{E}Y_k = 0$.

Considering the spectral norm $\|Y\|$ defined by

$$\|Y\| = \sup_{\|f_1\|\leq 1, \|f_2\|\leq 1} \langle f_1, Yf_2 \rangle = \sup_{\|f_1\|\leq 1, \|f_2\|\leq 1} \sum_{k=1}^{n} \langle f_1, Y_k f_2 \rangle \qquad (3.26)$$

The bound of $f(Y_k)$ can be estimated by

$$f(Y_k) \leq \left| \left\langle f_1, \frac{\delta_k u_k \otimes u_k - \mathrm{E}\delta_k u_k \otimes u_k}{m} f_2 \right\rangle \right|$$

$$\leq \frac{1}{m} |\langle f_1, u_k \rangle|^2 \leq \frac{1}{m} v^2 := B$$

for all $k$ with the probability exceeding $1-\delta$.

Now we can deal with the bound of $\mathrm{E}(f^2(Y_k))$ as

$$\mathrm{E}(f^2(Y_k)) \leq \mathrm{E}\left| \left\langle f_1, \frac{\delta_k u_k \otimes u_k}{m} f_2 \right\rangle \right|^2$$

$$= \frac{1}{mn} \mathrm{E}|\langle f, u_k \rangle|^4 \qquad (3.27)$$

It is noted that $\langle f, u_k \rangle$ is a random variable with zero-man and unit-variance; therefore, the value of $\mathrm{E}|\langle f, u_k \rangle|^4$ is the forth-moment of Gaussian random variable, i.e.

$$\sigma_g^4 = \mathrm{E}|\langle f, u_k \rangle|^4 = 3.$$

Now we can prove that

$$\sum_{k=1}^{n} \mathrm{E}(f^2(Y_k)) \leq \frac{1}{m} \sigma_g^4 = \frac{3}{m}.$$

In conclusion, applying the well-known Talagrand's inequality will yield to

$$\Pr(\|Y\| - E\|Y\| > t) \leq 3\exp\left(-\frac{t}{KB}\log\left(1 + \frac{Bt}{3/m + B\mathrm{E}\|Y\|}\right)\right)$$

$$= 3\exp\left(-\frac{mt}{Kv^2}\log\left(1 + \frac{t}{1 + \mathrm{E}\|Y\|}\right)\right)$$

(3.28)

Take $m$ enough large such that $E\|Y\| \leq \frac{1}{4}$ in above equation and pick $t = \frac{1}{4}$.

Now using $\log(1+x) > \frac{2}{3}x$ for $0 \leq x \leq 1$, we have

$$\Pr(\|Y\| > \tfrac{1}{2}) \leq 3\exp\left(-\frac{m}{30}\frac{1}{Kv^2}\right) \qquad (3.29)$$

which means $\Pr\left(\|Y\| > \tfrac{1}{2}\right) \leq \delta$ when $m \geq 30Kv^2 \log\left(\tfrac{3}{\delta}\right)$.

In addition, $\mathrm{E}\|Y\| \leq \dfrac{1}{4}$ if $m \geq 4C_R v\sqrt{\log|T|}$.

∎

**THEOREM 3.4.**

Let $U$ be a CS measurement matrix as above. Fix a subset $\Gamma$, let $r^k$ be the rows of $U_\Gamma$, and set $v := \max_{k=1,2,\cdots,n} \|r^k\|_2$. Choose a subset $\Omega$ of the measurement domain of size $|\Omega| = m$ and a sign sequence $z$ on $\Gamma$ uniformly at random. Set $\Phi = R_\Omega U$, the matrix constructed from the rows of $U$ indexed by . Suppose that

$$m \geq C_0 v^2 \log\left(n/\delta\right)$$

and also $m \geq C_0' \mu^2 \log^2\left(n/\delta\right)$, where $C_0$ and $C_0'$ are known constants. Then with probability exceeding $1 - O(\delta)$, every vector $\alpha_0$ supported on $\Gamma$ with sign sequence $z$ can be recovered from $y = \Phi \alpha_0$ by solving (1.2)

**LEMMA 3.5.**

Let $\Phi$, $\mu$, $\Gamma$ $v$ and $m$ be as in Theorem 3.3. Fix $\gamma \in \Gamma^c$, and consider the random vector $\upsilon_\gamma = \Phi_\Gamma^* \varphi_\gamma$. Assume that $\sqrt{m} \leq v^2/\mu$, Then for any $a^2 \leq \dfrac{\sqrt{m}}{\mu}$,

$$\Pr\left(\|\Phi_\Gamma^* \varphi_\gamma\|_2 \geq v\sqrt{m} + am^{1/4}\mu^{1/2}v\right) \leq 3\exp(-Ca^2) \qquad (3.40)$$

where $C$ is a known constant.

*Proof.*

This proof is also the application of the Talagrand' inequality and is carried out along the line carried out by Romberg in the following. Using the Bernoulli sampling model, we can write $\Phi_\Gamma^* \varphi_\gamma$ as a sum of independent random variables,

$$\Phi_\Gamma^* \varphi_\gamma = \sum_{k=1}^n \iota_k U_{k,\gamma} r^k \qquad (3.41)$$

where $r^k$ is the $k$th row of $U_\Gamma = H\Psi_\Gamma$. Note that $\mathrm{E}\left(\Phi_\Gamma^* \varphi_\gamma\right) = 0$. To bound the expected value of

$\|\Phi_\Gamma^* \varphi_\gamma\|_2$, we use

$$E\left\{\|\Phi_\Gamma^* \varphi_\gamma\|_2\right\}^2 \leq E\left\{\|\Phi_\Gamma^* \varphi_\gamma\|_2^2\right\}$$
$$= \sum_{k_1,k_2=1}^{n} E\left\{\iota_{k_1}\iota_{k_2} U_{k_1,\gamma} U_{k_2,\gamma}^* r^{k_1}\left(r^{k_2}\right)^*\right\} \quad (3.42)$$

By using

$$E\left\{U_{k_1,\gamma} U_{k_2,\gamma}^* r^{k_1}\left(r^{k_2}\right)^*\right\} = E\left(U_{k_1,\gamma} U_{k_2,\gamma}^*\right) E\left\{r^{k_1}\left(r^{k_2}\right)^*\right\}$$
$$+ E\left(U_{k_1,\gamma} r^{k_1}\right) E\left\{U_{k_2,\gamma}^* \left(r^{k_2}\right)^*\right\} + E\left(U_{k_1,\gamma} \left(r^{k_2}\right)^*\right) E\left\{U_{k_2,\gamma}^* r^{k_1}\right\} \quad (3.43)$$

one has

$$E\left\{\|\Phi_\Gamma^* \varphi_\gamma\|_2^2\right\} = \frac{m}{n}\sum_{k=1}^{n} E\left|U_{k,\gamma}\right|^2 \cdot E\|r^k\|^2 + \sum_{k_1,k_2=1}^{n} E\left(\iota_{k_1}\iota_{k_2}\right) E\left(U_{k_1,\gamma}\left(r^{k_2}\right)^*\right) E\left(U_{k_2,\gamma}^* r^{k_1}\right)$$
$$\leq \frac{mv^2}{n}\sum_{k=1}^{n} E\left|U_{k,\gamma}\right|^2 + 2n\left(\frac{m}{n}\right)^2$$
$$\approx mv^2 \quad (3.44)$$

It is noted that the fact has been employed that for the Gaussian random variables, the expected value of odd joint variables is zero while the mean of even joint variables is the sum of all possible combination.

For $\bar{\sigma}^2$, note that

$$\bar{\sigma}^2 = \sup_f \sum_{k=1}^{n} E\left|\langle f, Y_k \rangle\right|^2$$
$$= \frac{m}{n}\sup_f \sum_{k=1}^{n} E\left|U_{k,\gamma}\right|^2 \left|\langle f, r^k \rangle\right|^2 \quad (3.45)$$
$$= \frac{m}{n}\sup_f \sum_{k=1}^{n} E\left|\langle f, r^k \rangle\right|^2$$
$$= m \leq m\mu^2$$

where $E\left|U_{k,\gamma}\right|^2 = 1$ and $\sum_{k=1}^{n} E\left|\langle f, r^k \rangle\right|^2 = n$ have been used.

Note that for all $f \in \mathbb{R}^S$ with $\|f\| \leq 1$,

$$\left|\langle f, Y_k \rangle\right| = \left|\langle f, \iota_k U_{k,\gamma} r^k \rangle\right| \leq \left|U_{k,\gamma}\right|\left|\langle f, r^k \rangle\right| \leq \mu v := B \quad \text{for all } k$$

with probability exceeding $1-O(\delta)$. Plugging the bounds for $\mathrm{E}\|\Phi_\Gamma^* \varphi_\gamma\|_2$, B, and $\bar{\sigma}^2$ into Talagrand inequality, we have

$$\Pr\left(\|\Phi_\Gamma^* \varphi_\gamma\|_2 \geq v\sqrt{m} + t\right) \leq 3\exp\left(-\frac{t}{KB}\log\left(1 + \frac{Bt}{m\mu^2 + \mu v^2\sqrt{m}}\right)\right) \quad (3.46)$$

Using the fact that $\log(1+x) \geq 2x/3$ for $0 \leq x \leq 1$, this becomes

$$\Pr\left(\|\Phi_\Gamma^* \varphi_\gamma\|_2 \geq v\sqrt{m} + t\right) \leq 3\exp\left(-\frac{2t}{3KB}\frac{Bt}{m\mu^2 + \mu v^2\sqrt{m}}\right) \quad (3.47)$$

for all $0 \leq t \leq m\mu^2/B + v\sqrt{m} \leq 2v\sqrt{m}$. Thus

$$\Pr\left(\|\Phi_\Gamma^* \varphi_\gamma\|_2 \geq v\sqrt{m} + a\mu^{1/2}m^{1/4}v\right) \leq 3\exp(-Ca^2) \quad (3.48)$$

where $a^2 \leq 4\sqrt{m}/\mu$ and $C = \frac{1}{3K}$. Now, we have

$$\Pr\left(\|\Phi_\Gamma^* \varphi_\gamma\|_2 \geq v\sqrt{m} + a\mu^{1/2}m^{1/4}v\right) \leq 3\exp(-Ca^2) \quad (3.49)$$

∎

To finish off the proof of the Theorem, let $\mathcal{A}$ be the event that Eq. () holds; step 1 tell us that $\Pr(\mathcal{A}) \leq \delta$. Let $\mathcal{B}_\lambda$ be the event that

$$\max_{\gamma \in \Gamma^c}\left\|(\Phi_\Gamma^* \Phi_\Gamma)^{-1} \upsilon_\gamma\right\|_2 \leq \lambda \quad (3.50)$$

where $\lambda = 2vm^{-1/2} + 2a\mu^{1/2}m^{-3/4}v$. By Lemma 3.5 and taking the union bound over all $\gamma \in \Gamma^c$, we have

$$\Pr(\mathcal{B}_\lambda | \mathcal{A}) \leq 3n\exp(-Ca^2) \quad (3.51)$$

By the Hoeffding inequality,

$$\Pr\left(\max_{\gamma \in \Gamma^c}|\pi(\gamma)| > \alpha | \mathcal{B}_\lambda, \mathcal{A}\right) \leq 2n\exp\left(-\alpha^2/2\lambda^2\right) \quad (3.52)$$

Our final probability of success can then be bounded by

$$\Pr\left(\max_{\gamma \in T^c}|\pi(\gamma)| > \alpha\right) \leq \Pr\left(\max_{\gamma \in T^c}|\pi(\gamma)| > \alpha | \mathcal{B}_\lambda, \mathcal{A}\right) + \Pr(\mathcal{B}_\lambda | \mathcal{A}) + \Pr(\mathcal{A})$$
$$\leq 2n\exp\left(-\alpha^2/2\lambda^2\right) + 3n\exp(-Ca^2) + \delta$$

(3.53)

Then we can make the second term in above expression less than $\delta$ by choosing

$$a = C^{-1/2}\sqrt{\log(3n/\delta)}; \qquad (3.54)$$

Because of $a^2 \leq \frac{\sqrt{m}}{\mu}$, one has $m \geq 16C^{-2}\mu^2 \log^2(3n/\delta)$. This choice of $a$ also ensures that

$$\lambda \leq 3vm^{-1/2}$$

For the first term in above equation to be less than $\delta$, we need

$$\lambda^2 \leq \frac{\alpha^2}{2\log(2n/\delta)}$$

which hold when

$$m \geq \frac{18}{\alpha^2} v^2 \log(2n/\delta)$$

Especially, selecting $\alpha = 1/2$, $m \geq 72v^2 \log(2n/\delta)$. Noting that $v^2 \leq 8T$ with probability exceeding $1-\delta$, hence, $m \geq CT\log(2n/\delta)$ with probability exceeding $1-\delta$.

## IV. CONCLUSIONS

In this paper, the theoretical analysis of compressive sensing via random filter, firstly outlined by J. Romberg [compressive sensing by random convolution, *submitted to SIAM Journal on Imaging Science on July 9, 2008*], has been refined or generalized to the design of general random filter used for compressive sensing. Theorems 2.1 and 1.2 tell us that we can recover perfectly a *S* sparse signal from on the order of *S*log*n*. If we are willing to pay additional log factors, we can also guarantee that the recovery will stable.

ACKNOWLDGEMENT:
This work has been supported by the National Natural Science Foundation of China under Grants 60701010 and 40774093.